\documentstyle[12pt]{article}

\begin{document}
\parskip 1ex
\setcounter{page}{1}
\oddsidemargin 0pt
\evensidemargin 0pt
\topmargin -40pt
%
\newcommand{\be}{\begin{equation}}
\newcommand{\ee}{\end{equation}}
\newcommand{\beq}{\begin{eqnarray}}
\newcommand{\eeq}{\end{eqnarray}}
\def\a{\alpha}
\def\b{\beta}
\def\g{\gamma}
\def\G{\Gamma}
\def\d{\delta}
\def\e{\epsilon}
\def\z{\zeta}
\def\h{\eta}
\def\th{\theta}
\def\k{\kappa}
\def\l{\lambda}
\def\L{\Lambda}
\def\m{\mu}
\def\n{\nu}
\def\x{\xi}
\def\X{\Xi}
\def\p{\pi}
\def\P{\Pi}
\def\r{\rho}
\def\s{\sigma}
\def\S{\Sigma}
\def\t{\tau}
\def\f{\phi}
\def\F{\Phi}
\def\c{\chi}
\def\w{\omega}
\def\W{\Omega}
\def\de{\partial}

\def\pct#1{(see Fig. #1.)}

\begin{titlepage}
\hbox{\hskip 12cm ROM2F/2001/01  \hfil}
\hbox{\hskip 12cm IHP-2000/10 \hfil}
\hbox{\hskip 12cm hep-th/0101074 \hfil}
\vskip 1.4cm
\begin{center}  {\Large  \bf All \ Couplings \ of 
\vskip .5cm 
Minimal \ Six-dimensional \ Supergravity}
\vspace{1.5cm}
 
{\large \large  Fabio Riccioni}
\vspace{0.6cm}

{\sl Dipartimento di Fisica, \ \ Universit{\`a} di Roma \ ``Tor Vergata'' \\
I.N.F.N.\ - \ Sezione di Roma \ ``Tor Vergata'' \\ Via della Ricerca
Scientifica, 1 \ \ \ 00133 \ Roma \ \ Italy \\
and \\ 
Centre \ Emile \ Borel, \ Institut \ Henri \ Poincar\'e \\
11 rue Pierre et Marie Curie, \ F-75321 \ Paris \ Cedex 05 \ \ France}
\end{center}
\vskip 1.5cm

\abstract{We describe the complete coupling of $(1,0)$ six-dimensional 
supergravity to tensor, vector and hypermultiplets. 
The generalized Green-Schwarz mechanism implies that the resulting 
theory embodies factorized gauge and supersymmetry 
anomalies, to be disposed of by fermion loops. Consequently, the low-energy 
theory is determined by the Wess-Zumino 
consistency conditions, rather than by the requirement of supersymmetry. 
As already shown for the case without 
hypermultiplets, this procedure does not fix 
a quartic coupling for the gauginos. 
With respect to these previous results, the inclusion of charged hypermultiplets
gives additional terms in the supersymmetry anomaly. We also consider the case 
in which abelian vectors are present. As in 
the absence of hypermultiplets, abelian vectors allow additional couplings.
Finally, we apply the 
Pasti-Sorokin-Tonin prescription to this model.}

\vskip 1.5cm
\begin{center}
{( January , \ 2001 )}
\end{center}
\vfill
\end{titlepage}
\makeatletter
\@addtoreset{equation}{section}
\makeatother
\renewcommand{\theequation}{\thesection.\arabic{equation}}
\addtolength{\baselineskip}{0.3\baselineskip} 

\section{Introduction}
Perturbative six-dimensional string vacua with minimal supersymmetry can arise 
for instance
as compactifications of the heterotic string on $K3$, or as parameter-space 
orbifolds (orientifolds) \cite{cargese} of $K3$ reductions of the type-IIB string.
While in the former case only a single tensor multiplet is present,
in the latter one obtains vacua with variable numbers 
of tensor multiplets \cite{bs},
related by string dualities to non-perturbative heterotic and M-theory vacua.
In these models, the anomalous contribution due to fermion loops is derived from 
the residual anomaly polynomial  
$$
\ c^r_x \ c^s_y \ \eta_{rs} \ {\rm tr}_x F^2 \ {\rm tr}_y F^2
\quad ,
$$
where the $c$'s are a collection of constants ($x$ and $y$ run over the various
semi-simple Lie factors in the gauge group and over the Lorentz group) 
and $\eta$ is the Minkowski metric for
$SO(1,n_T)$, with $n_T$ the number of tensor multiplets \cite{as}.
As a consequence, several antisymmetric
tensors take part in a generalized Green-Schwarz mechanism \cite{gs,as}.
The corresponding Green-Schwarz term has the form 
$$
B^r \ c_r^x \ {\rm tr}_x F^2 
$$
and, if one considers only gauge anomalies, 
contains only two derivatives, and thus belongs 
to the low-energy effective action.
Consequently, the resulting low-energy 
lagrangian has a ``classical'' gauge anomaly, that the 
Wess-Zumino  conditions \cite{wz} relate to a ``classical'' supersymmetry
anomaly \cite{fms}. 

The complete coupling of (1,0) six-dimensional supergravity
to non-abelian vector and tensor multiplets, obtained in \cite{frs} requiring 
the closure of the Wess-Zumino conditions, has revealed another related 
aspect of these six-dimensional models: a quartic coupling 
for the gauginos is undetermined, and the construction is consistent for any 
choice of this coupling. Correspondingly, the commutator of two supersymmetry
transformations on the gauginos contains an extension, that plays a 
crucial role in ensuring that the Wess-Zumino consistency conditions close 
on-shell.
The coupling of (1,0) six-dimensional supergravity to non-abelian vectors and 
self-dual tensors reveals neatly 
the realization of a peculiar aspect of the 
physics of branes: singularities in the gauge couplings appear for particular
values of the scalars in the tensor multiplets \cite{as}, and can be ascribed to 
a phase transition \cite{dmw} in which a string  becomes tensionless 
\cite{tensionless}. 
Moreover, as was shown in \cite{rs}, in this model 
the divergence of the energy-momentum 
tensor is non-vanishing, as is properly the case for a theory that 
has  gauge anomalies but no gravitational anomalies (gravitational anomalies 
could be accounted for introducing  higher-derivative couplings).
The whole construction can also be repeated with 
the inclusion of abelian vectors, that actually allow more general 
couplings, since in this case the residual 
anomaly polynomial can have the more general form 
$$
\ c^r_{ab} \ c^s_{cd} \ \eta_{rs} F^a \wedge F^b \wedge F^c \wedge F^d
\quad ,
$$
where the indices $a,b,c,d$ run over the different $U(1)$ gauge groups,
and where the $c^r$'s are symmetric matrices that may not be simultaneously 
diagonalized \cite{r}. 

Notice that these low-energy couplings are obtained by consistency once one 
includes the Green-Schwarz term in the low-energy theory. 
The complete theory, supersymmetric and gauge-invariant,
would also include additional non-local couplings arising from fermion loops.   
This is exactly as in the ten-dimensional case, what is peculiar of 
these six-dimensional models is that here the anomalous terms belong 
to the low-energy effective action.

In order to have an explicit realization of the low-energy dynamics 
of six-dimensional string vacua, it is of interest to consider 
how the whole construction is modified by the inclusion 
of hypermultiplets.  
In \cite{ns}, the complete coupling to a single tensor multiplet and to vector and 
charged hypermultiplets was obtained for the case in which no anomalies and no
singular couplings are present. 
More recently, an analysis of the case in which various tensor multiplets
are present was carried out in \cite{ns1}, however 
without taking into account the
anomalous terms. Still, this analysis shows that, in correspondence to the phase 
transition, additional singular terms appear because of the presence 
of charged hypermultiplets.

In this paper we construct the complete coupling of (1,0) supergravity 
to all possible (1,0) multiplets, generalizing the results of 
\cite{frs} in order to include hypermultiplets, 
and extending the results of \cite{ns1} to all orders in the 
fermi fields, while taking into account the anomalous couplings. 
We show that the inclusion of charged hypermultiplets gives 
additional terms in the supersymmetry anomaly. As was the case without 
hypermultiplets \cite{frs}, the resulting theory is determined up 
to a quartic coupling for the gauginos, and correspondingly the supersymmetry 
algebra contains an extension that guarantees the consistency
of the construction. Following \cite{ns,ns1}, we will consider the case 
in which the scalars in the hypermultiplets parametrize the coset 
$USp(2,2n_H)/USp(2) \times USp(2n_H)$, and we will describe the gauging of the 
full compact subgroup $USp(2) \times USp(2 n_H)$ 
of the isometry group $USp(2,2n_H)$. 
Other cases, in which the scalars parametrize more general 
quaternionic symmetric spaces or are charged with respect to different 
subgroups of the isometry group, 
can be straightforwardly obtained from our results.

The paper is organized as follows. In section 2 we construct the complete 
(1,0) supergravity coupled to $n_T$ tensor multiplets, non-abelian vector 
multiplets and $n_H$ hypermultiplets. In 
section 3 we describe
the case in which abelian vectors are included. Section 4 is devoted 
to a discussion, in which we also show how to apply the 
Pasti-Sorokin-Tonin (PST) construction \cite{pst} to this model. Finally,   
the appendix collects some details on the notation and some useful identities. 

\section{Supersymmetry algebra and equations of motion}
In this section we describe the full coupling of six-dimensional supergravity 
to vector, tensor and hypermultiplets. We will use notations similar to the 
ones of  
\cite{frs} for what concerns the coupling to vector and tensor multiplets, while 
in the description of the coupling to hypermultiplets we will follow 
the notation of 
\cite{ns}. Some details about our conventions are contained in the appendix. 
 
We first summarize the field content of the theory. The gravitational 
multiplet contains the vielbein $e_\m{}^m$, a 2-form and 
a left-handed gravitino $\psi_\m^A$, the tensor multiplet contains 
a 2-form, a scalar and a right-handed tensorino, the  
vector multiplet contains a vector $A_\m$ and a left-handed gaugino $\l^A$, 
and finally the 
hypermultiplet contains four scalars and a right-handed hyperino. 
In the presence of $n_T$ tensor multiplets,
the tensorinos are denoted by 
$\chi^{MA}$ where $M=1,...,n_T$ is an $SO(n_T )$ index. The index $A=1,2$
is in the fundamental representation of $USp(2)$, and the gravitino, the 
tensorinos and the gauginos are $USp(2)$ doublets satisfying 
the symplectic-Majorana condition
\be
\psi^A =\e^{AB} C \bar{\psi}^T_B \quad. 
\ee
The $n_T$ scalars in the tensor multiplets 
parametrize the coset 
$SO(1,n_T)/SO(n_T)$, while the $(n_T +1)$ 2-forms from the gravitational 
and tensor multiplets are collectively 
denoted by $B^r_{\m\n}$,
with $r=0,...,n_T$ in the fundamental representation of $SO(1,n_T)$, 
and their field-strengths satisfy 
(anti)self-duality conditions.
The vector and the gaugino are in the adjoint representation of the gauge group.
Finally, taking into account $n_H$ hypermultiplets,
the hyperinos are denoted by $\Psi^{a}$, where $a=1,...,2n_H$ is a 
$USp(2n_H)$ index, and the symplectic-Majorana condition for these spinors
is
\be
\Psi^a =\W^{ab} C \bar{\Psi}^T_b \quad,
\ee
where $\W^{ab}$ is the 
antisymmetric invariant tensor of $USp(2n_H)$ (see the appendix for more details). 
The hyper-scalars $\phi^\a$, $\a=1,...,4n_H$, are coordinates of a quaternionic
manifold, that is a manifold whose holonomy group is contained in 
$USp(2) \times USp(2n_H)$.

If the quaternionic manifold parametrized 
by the hyper-scalars has isometries, 
these correspond to global symmetries of the supergravity theory. 
Then the global symmetry group, or a subgroup thereof, can be gauged. 
Following \cite{ns,ns1}, 
we will consider
without loss of generality the case in which the scalars parametrize the 
symmetric manifold $USp(2,2n_H)/USp(2)\times USp(2n_H)$, 
whose isometry group is $USp(2,2 n_H )$. We will then describe 
the gauging of the maximal compact subgroup $USp(2)
\times USp(2n_H )$ of the isometry group. All the results can be  
naturally generalized to other symmetric quaternionic spaces
\footnote{I am grateful to S. Ferrara for discussions about this point.}. 

The scalars in the tensor multiplets 
can be described, following \cite{romans}, in terms of 
the $SO(1,n_T)$ matrix
\be
V = \pmatrix{v_r \cr x^M_r} \quad ,
\ee
whose elements satisfy the constraints
\be
v^r v_r =1 \quad , \qquad  v_r v_s - x^M_r x^M_s = \eta_{rs}\quad , \qquad 
v^r x^M_r =0 \quad . \label{scalars}
\ee
In the following, we will 
take $v_r$ and $x^M_r$, with the constraints of eq. (\ref{scalars}), 
as fundamental fields, as in \cite{romans}, so that the composite $SO(n_T)$
connection that appears in the covariant derivative of the tensorinos will be
$x^N_r \de_\m x^{Mr}$. On the other hand, the notation of \cite{ns1}, 
in which the fundamental fields are the scalars $\Phi^{\bar{\a}}$ ($\bar{\a}
=1,...,n_T $)
parametrizing the coset manifold,  adds to 
the supersymmetry variation of the tensorinos 
$\chi^{MA}$  the term
\be
-\delta \Phi^{\bar{\a}} {\cal{A}}_{\bar{\a}}^{MN} \chi^{NA} \quad ,
\ee
where ${\cal{A}}_{\bar{\a}}^{MN}$ is the composite connection 
of $SO(n_T)$ \cite{ns1}. 
In this notation, the commutator of two supersymmetry transformations on the
tensorinos does not generate a local $SO(n_T)$ transformation.

We now recall the notations used to describe 
the scalars in the hypermultiplets.
We denote by $V_\a^{aA}(\phi )$ the vielbein of the quaternionic manifold, 
where the index structure corresponds to the requirement that the 
holomony be contained
in $USp(2)\times USp(2n_H )$. The internal $USp(2)$ and
$USp(2n_H)$ connections are then denoted, respectively, by 
${\cal{A}}_\a^A{}_B$  and ${\cal{A}}_\a^a{}_b$, that in our conventions 
are anti-hermitian matrices. The index $\a =1,...,4 n_H$ is a curved 
index on the quaternionic manifold. 
The field-strengths of the connections are
\beq
& & {\cal{F}}_{\a\b}{}^A{}_B = \de_{\a} {\cal{A}}_\b^A{}_B -\de_{\b} 
{\cal{A}}_\a^A{}_B
+[ {\cal{A}}_\a , {\cal{A}}_\b ]^A{}_B \quad ,\nonumber\\
& & {\cal{F}}_{\a\b}{}^a{}_b = \de_{\a} {\cal{A}}_\b^a{}_b -\de_{\b} 
{\cal{A}}_\a^a{}_b
+[ {\cal{A}}_\a , {\cal{A}}_\b ]^a{}_b \quad ,
\eeq
where $\de_\a = \de / \de \phi^\a $. 
The request that the vielbein $V_\a^{aA}(\phi )$ 
be covariantly constant gives the following relations \cite{bw}:
\beq
& & V^{\a}_{aA} V^{\b}_{bB}g_{\a\b}=\W_{ab} \e_{AB} \quad ,\nonumber\\
& & V^{\a}_{aA} V^{\b bA} + V^{\b}_{aA} V^{\a bA} =\frac{1}{n_{H}}
g^{\a\b}\delta^a_b 
\quad , \nonumber\\
& & V^{\a}_{aA} V^{\b aB} +V^{\b}_{aA} V^{\a aB}=g^{\a\b}\delta^A_B \quad ,
\eeq 
where $\W_{ab}$ is the antisymmetric invariant tensor of $USp(2n_H)$.
The raising and lowering conventions are collected in the appendix. 
The field-strength of the $USp(2)$ connection ${\cal{A}}_\a^A{}_B$  
is naturally constructed in terms of $V_{\a}^{aA}$ by the relation: 
\be
{\cal{F}}_{\a\b AB}= V_{\a aA}V_\b^a{}_B +V_{\a aB}V_\b^a{}_A \quad ,
\ee
and then the cyclic identity for the internal curvature tensor implies 
that the field-strength of
the $USp(2n_H)$ connection ${\cal{A}}_\a^a{}_b$ has the form
\be
{\cal{F}}_{\a\b ab}= V_{\a aA}V_{\b b}{}^A +V_{\a bA}V_{\b a}{}^A  + \W_{abcd}
V_\a^{dA}V_\b^c{}_A \quad , 
\ee 
where $\W_{abcd}$ is totally symmetric in its indices \cite{bw}.

Now, assuming that the scalars parametrize the coset manifold 
$USp(2, 2n_H ) / USp(2)\times USp(2n_H)$, 
we describe the gauging of the hypermultiplets under 
the group $USp(2)\times USp(2n_H)$ \cite{ns}, 
that is the maximal compact subgroup of the isometry group. 
We denote the gauge fields of this group by $A_\m^i$ and $A_\m^I$,
where $i$ and $I$ take values in the adjoint representation of $USp(2)$ and 
$USp(2n_H )$, and the corresponding field-strengths are
\beq
& & F_{\m\n}^i =\de_\m A_\n^i -\de_\n A_\m^i + \e^{ijk} A_\m^j A_\n^k \quad,
\nonumber\\
& & F_{\m\n}^I =\de_\m A_\n^I -\de_\n A_\m^I + f^{IJK} A_\m^J A_\n^K \quad,
\eeq
where $\e^{ijk}$ and $f^{IJK}$ are the structure constants of $USp(2)$
and $USp(2n_H )$.
Under the gauge transformations 
\be
\delta A_\m^i = D_\m \L^i \quad , \qquad \delta A_\m^I =D_\m \L^I 
\ee
the scalars transform as
\be
\delta \phi^\a = \L^i \xi^{\a i} + \L^I \xi^{\a I}\quad ,
\ee
where $\xi^{\a i}$ and $\xi^{\a I}$ are the Killing vectors corresponding to
the $USp(2)$ and $USp(2n_H )$ isometries.
The covariant derivative for the scalars is then
\be
D_\m \phi^\a =\de_\m \phi^\a - A_\m^i \xi^{\a i} -A_\m^I \xi^{\a I} \quad .
\ee

One can correspondingly define the covariant derivatives for the spinors in 
a natural way, adding the composite connections
$D_\m \phi^\a {\cal{A}}_\a$. 
For instance, the covariant derivative for the hyperinos 
$\Psi^a$ will contain the connections $D_\m \phi^\a {\cal{A}}_\a^a{}_b$, 
while the covariant derivative for the gravitino and the tensorinos 
will contain the connections
$D_\m \phi^\a {\cal{A}}_\a^A{}_B$. The covariant derivatives for the 
gauginos $\l^{iA},\l^{IA}$ are 
\beq
D_\m \l^{iA}=\de_\m \l^{iA} +\frac{1}{4}\w_{\m mn}\g^{mn}\l^{iA}
+ D_\m \phi^\a {\cal{A}}_\a^A{}_B \l^{i B}+\e^{ijk} A_\m^j \l^{kA}
\quad ,\nonumber\\
D_\m \l^{IA}=\de_\m \l^{IA} +\frac{1}{4}\w_{\m mn}\g^{mn}\l^{IA}
+ D_\m \phi^\a {\cal{A}}_\a^A{}_B \l^{I B}+ f^{IJK} A_\m^J \l^{KA}
\quad .
\eeq
Notice that the gravitino, the tensorinos and the hyperinos 
are not coupled to the gauge vectors through terms
that do not contain the hyper-scalars. 

We now proceed to the construction of the model.
We assume that the gauge group has the form 
$G= \prod_z G_z$, with $G_z$ semi-simple. The scalars in the hypermultiplets
are charged with respect to   
$G_1 =USp(2)$ and $G_2 =USp(2n_H)$. The field-strengths of the 2-forms
$B^r_{\m\n}$ are
\be
H^r_{\m\n\r}= 3\de_{[\m} B^r_{\n\r ]} + c^{r z} \w^z_{\m\n\r} \quad,
\ee
where $c^{rz}$ are constants and $\w^z$ are 
the Chern-Simons 3-forms:
\be 
\w^z = tr_z (A dA +\frac{2}{3} A^3 ) \quad .
\ee 
These 3-form field-strengths satisfy 
(anti)self-duality conditions, that to lowest order in the fermi fields are
\be
G_{rs} H^{s \m\n\r} = \frac{1}{6e} \e^{\m\n\r\s\delta \tau} H_{r \s \delta\tau}
\quad ,
\ee
where $G_{rs}=v_r v_s + x^M_r x^M_s$.
Gauge invariance of $H^r$ requires that $B^r $ transform under vector gauge
transformations according to
\be
\d B^r =- c^{rz} tr_z (\L dA ) \quad .
\ee

To lowest order in the fermi fields, we reproduce the construction of \cite{fms}, 
adding the hypermultiplet couplings. The equations for all 
fields, with the exception of the 2-forms, can be obtained from the lagrangian
\beq
e^{-1}{\cal{L}} & & =-\frac{1}{4}R +\frac{1}{12}G_{rs} H^{r \m\n\r} H^s_{\m\n\r}
-\frac{1}{4} \de_\m v^r \de^\m v_r 
+\frac{1}{2} v_r c^{rz} tr_z (F_{\m\n}F^{\m\n}) \nonumber\\
& & +\frac{1}{8e}\e^{\m\n\r\sigma\delta\tau}B^r_{\m\n}c_r^z tr_{z}(F_{\r\sigma}
F_{\delta\tau})
+\frac{1}{2} g_{\a\b}(\phi ) D_\m \phi^\a D^\m \phi^\b\nonumber \\
& & 
+\frac{1}{4v_r c^{r1}} {\cal{A}}_\a^A{}_B 
{\cal{A}}_\b^B{}_A \xi^{\a i} \xi^{\b i} 
+\frac{1}{4v_r c^{r2}} {\cal{A}}_\a^A{}_B 
{\cal{A}}_\b^B{}_A \xi^{\a I} \xi^{\b I}
\nonumber \\
& & -\frac{i}{2}(\bar{\psi}_\m \g^{\m\n\r} D_\n 
\psi_\r )-\frac{i}{2}v_r H^{r \m\n\r}(\bar{\psi}_\m \g_\n \psi_\r)
+\frac{i}{2} (\bar{\chi}^M \g^\m D_\m \chi^M )\nonumber \\ 
& & -\frac{i}{24}v_r H^r_{\m\n\r} (\bar{\chi}^M \g^{\m\n\r}
\chi^M ) +\frac{1}{2}x^M_r \de_\n v^r (\bar{\psi}_\m \g^\n
\g^\m \chi^M) -\frac{1}{2} x^M_r H^{r \m\n\r} ( \bar{\psi}_\m
\g_{\n\r} \chi^M )\nonumber\\
& & +\frac{i}{2}(\bar{\Psi}_a \g^\m D_\m \Psi^a ) +\frac{i}{24}
v_r H^r_{\m\n\r} (\bar{\Psi}_a \g^{\m\n\r}\Psi^a ) - V_\a^{aA}D_\n \phi^\a
(\bar{\psi}_{\m A} \g^\n \g^\m \Psi_a )\nonumber \\
& & -i v_r c^{rz} tr_z(\bar{\l} \g^\m D_\m \l )
-\frac{i}{\sqrt{2}} v_r c^{rz} tr_z[F_{\n\r} (\bar{\psi}_\m \g^{\n\r}\g^\m \l )]
\nonumber\\
& & -\frac{1}{\sqrt{2}} x^M_r c^{rz} tr_z [F_{\m\n} (\bar{\chi}^M \g^{\m\n}\l )]
+\frac{i}{12}c_r^z H^r_{\m\n\r} tr_z (\bar{\l} \g^{\m\n\r} \l)
\nonumber\\
& & -\sqrt{2} V_\a^{aA} [\xi^{\a i}(\bar{\l}^i_A \Psi_a ) +\xi^{\a I} (\bar{\l}^I_A
\Psi_a )] +\frac{i}{\sqrt{2}}{\cal{A}}_\a^A{}_B 
[\xi^{\a i }(\bar{\l}^i_A \g^\m \psi_\m^B )
+ \xi^{\a I} (\bar{\l}^I_A \g^\m \psi_\m^B )] \nonumber \\
& & +\frac{1}{\sqrt{2}}{\cal{A}}_\a^A{}_B 
[ \frac{x^M_r c^{r1}}{v_s c^{s1}}\xi^{\a i}
(\bar{\l}^i_A \chi^{MB} ) + \frac{x^M_r c^{r2}}{v_s c^{s2}}\xi^{\a I}
(\bar{\l}^I_A \chi^{MB} ) ]\quad,
\label{lag}
\eeq
after imposing the (anti)self-duality conditions. 
With this prescription, its variation under the supersymmetry transformations
\beq 
& & \delta e_\m{}^m = -i ( \bar{\e} \g^m \psi_\m ) \quad , \nonumber\\ 
& & \delta B^r_{\m\n} =i v^r ( \bar{\psi}_{[\m} \g_{\n]} \e )
+\frac{1}{2} x^{Mr} ( \bar{\chi}^M \g_{\m\n} \e ) 
+ 2c^{rz}tr_z( A_{[\m} \delta A_{\n]} )
\quad , \nonumber\\ 
& & \delta v_r = x^M_r ( \bar{\e} \chi^M ) \quad , \qquad \delta x^M_r = v_r
(\bar{\e} \chi^M )\quad , \nonumber\\
& & \delta \phi^\a = V^\a_{aA} ({\bar{\e}}^A \Psi^a ) \quad ,\nonumber\\
& & \delta A_\m = -\frac{i}{\sqrt{2}} (\bar{\e} \g_\m \l )\quad, 
\nonumber\\
& & \delta \psi_\m^A = D_\m \e^A +\frac{1}{4} v_r H^r_{\m\n\r} \g^{\n\r} \e^A 
\quad , \nonumber\\
& & \delta \chi^{MA} =\frac{i}{2} x^M_r
\de_\m v^r \g^\m \e^A +\frac{i}{12} x^M_r H^r_{\m\n\r} \g^{\m\n\r} 
\e^A \quad ,\nonumber\\
& & \delta \Psi^a = i \g^\m \e_A V_\a^{aA} D_\m \phi^\a \quad , \nonumber\\
& & \delta \l^A = -\frac{1}{2\sqrt{2}} F_{\m\n} \g^{\m\n} \e^A \qquad 
\quad (z \neq 1,2 ) \quad , \nonumber \\
& & \delta \l^{iA}= -\frac{1}{2\sqrt{2}} F^{i}_{\m\n} \g^{\m\n} \e^A 
-\frac{1}{\sqrt{2}v_r c^{r1}} {\cal{A}}_\a^A{}_B \xi^{\a i} \e^B \nonumber\\
& & \delta \l^{IA}= -\frac{1}{2\sqrt{2}} F^{I}_{\m\n} \g^{\m\n} \e^A 
-\frac{1}{\sqrt{2}v_r c^{r2}} {\cal{A}}_\a^A{}_B \xi^{\a I} \e^B 
\eeq
gives the supersymmetry anomaly
\be 
{\cal{A}}_\e =-\frac{1}{4} \e^{\m\n\r\sigma\delta\tau} c_r^z c^{r z^\prime} tr_z (
\delta_\e A_\m A_\n ) tr_{z^\prime} (F_{\r\sigma} F_{\delta\tau})  
-\frac{1}{6} \e^{\m\n\r\sigma\delta\tau}
c_r^z c^{r z^\prime} tr_z (
\delta_\e A_\m F_{\n\r} ) \w^{z^\prime}_{\sigma\delta\tau} \ ,
\label{susyanomaly}
\ee
related by the Wess-Zumino conditions to the consistent gauge anomaly
\be 
{\cal{A}}_\L =- \frac{1}{4} \e^{\m\n\r\sigma\delta\tau} c_r^z c^{rz^\prime} tr_z (\L
\de_\m A_\n ) tr_{z^\prime} (F_{\r\sigma} F_{\delta\tau} )\quad .
\label{consanomaly}
\ee
Notice the presence in the lagrangian of the scalar potential
\be
V(\phi )=-\frac{1}{4v_r c^{r1}} {\cal{A}}_\a^A{}_B 
{\cal{A}}_\b^B{}_A \xi^{\a i} \xi^{\b i} 
-\frac{1}{4v_r c^{r2}} {\cal{A}}_\a^A{}_B 
{\cal{A}}_\b^B{}_A \xi^{\a I} \xi^{\b I} \quad .
\label{pot}
\ee
As in rather more conventional gauged models, the potential contains 
interesting informations, and it may be very instructive to study its extrema 
in special cases.  

We now want to extend the results to all orders in the fermi fields. 
First of all, we define the supercovariant quantities
\beq
& & \hat{\w}_{\m\n\r} = \w^0_{\m\n\r}  -\frac{i}{2} (\bar{\psi}_\m \g_\n \psi_\r 
+\bar{\psi}_\n \g_\r
\psi_\m +\bar{\psi}_\n \g_\m \psi_\r )\quad ,\nonumber\\
& & \hat{H}^r_{\m\n\r} = H^r_{\m\n\r}  -\frac{1}{2} x^{Mr} ( \bar{\chi}^M \g_{\m\n}
\psi_\r +
\bar{\chi}^M 
\g_{\n\r}\psi_\m + \bar{\chi}^M \g_{\r\m} \psi_\n )  \nonumber\\ 
& & \quad \quad - \frac{i}{2} 
v^r (\bar{\psi}_\m \g_\n \psi_\r +\bar{\psi}_\n \g_\r
\psi_\m +\bar{\psi}_\r \g_\m \psi_\n ) \quad ,\nonumber\\
& & \hat{\de_\m v^r} = \de_\m v^r -x^{Mr} (\bar{\chi}^M \psi_\m )\quad ,\nonumber\\
& & \hat{D_\m \phi^\a } =D_\m \phi^a - V^\a_{aA} (\bar{\psi}_\m^A \Psi^a )\quad,
\nonumber\\
& & \hat{F}_{\m\n}= F_{\m\n}+\frac{i}{\sqrt{2}}(\bar{\l} \g_\m 
\psi_\n )-\frac{i}{\sqrt{2}}(\bar{\l} \g_\n \psi_\m )\quad ,
\eeq
and require that the transformation rules for the fermi fields be supercovariant.
All fermionic terms in the supersymmetry transformations of the fermi fields 
that are not determined by supercovariance are then obtained 
requiring the closure of the supersymmetry algebra 
on bose and fermi fields. 
Moreover, since the supersymmetry algebra on the fermi 
fields closes only on-shell, in this way one can determine the complete fermionic 
field equations, and from these the complete lagrangian, up to some subtleties
related to the (anti)self-dual forms, that will be described in section 4. 

The complete supersymmetry transformations of the fermi fields are
\beq
& & \delta \psi_\m^A =D_\m (\hat{\w})\e^A +\frac{1}{4} v_r \hat{H}^r_{\m\n\r}
\g^{\n\r}\e^{A} -\frac{3i}{8} \g_\m \chi^{MA} (\bar{\e} \chi^M ) -\frac{i}{8} 
\g^\n \chi^{MA}
(\bar{\e} \g_{\m\n} \chi^M )\nonumber\\
& & \quad \quad +\frac{i}{16} \g_{\m\n\r} \chi^{MA} (\bar{\e} 
\g^{\n\r} \chi^M ) 
+ \frac{9i}{8} v_r c^{rz} tr_z [\l^A (\bar{\e} \g_\m \l)] -  
\frac{i}{8} v_r c^{rz} tr_z [\g_{\m\n} \l^A (\bar{\e} \g^\n \l )]\nonumber\\
& & \quad \quad + \frac{i}{16} v_r c^{rz} tr_z [\g^{\n\r} \l^A (\bar{\e}
\g_{\m\n\r} \l )] -\delta \phi^\a {\cal{A}}_\a^A{}_B \psi_\m^B \quad ,\nonumber\\ 
& & \delta \chi^{MA} =
\frac{i}{2} x^M_r (\hat{\de_\m v^r} ) \g^\m \e^A +
\frac{i}{12} x^M_r \hat{H}^r_{\m\n\r} \g^{\m\n\r}\e^A \nonumber\\
& & \quad \quad -\frac{1}{2} x^M_r c^{rz} tr_z [ \g_\m \l^A (\bar{\e} \g^\m \l ) ] 
-\delta \phi^\a {\cal{A}}_\a^A{}_B \chi^{MB} \quad ,\nonumber\\ 
& & \delta \Psi^a = i \g^\m \e_A V_\a^{aA} \hat{D_\m \phi^\a} -\delta \phi^\a
{\cal{A}}_\a^a{}_b \Psi^b \quad , \nonumber \\
& & \delta \l^{A}
=-\frac{1}{2\sqrt{2}}\hat{F}_{\m\n} \g^{\m\n} \e^A 
- \frac{x^M_r c^{rz}}{2v_s c^{sz}} (\bar{\chi}^M \l ) \e^A 
- \frac{x^M_r c^{rz}}{4v_s c^{sz}} (\bar{\chi}^M \e ) \l^{A}  
\nonumber \\
& & \quad \quad + 
\frac{x^M_r c^{rz}}{8v_s c^{sz}} (\bar{\chi}^M \g_{\m\n} \e ) \g^{\m\n}
\l^{A} -\delta \phi^\a {\cal{A}}_\a^A{}_B \l^{B}
\qquad \qquad \qquad \quad (z \neq 1,2 )\quad , \nonumber\\
& & \delta \l^{iA}
=-\frac{1}{2\sqrt{2}}\hat{F}^i_{\m\n} \g^{\m\n} \e^A 
- \frac{x^M_r c^{r1}}{2v_s c^{s1}} (\bar{\chi}^M \l^i ) \e^A 
- \frac{x^M_r c^{r1}}{4v_s c^{s1}} (\bar{\chi}^M \e ) \l^{iA}  
\nonumber \\
& & \quad \quad + 
\frac{x^M_r c^{r1}}{8v_s c^{s1}} (\bar{\chi}^M \g_{\m\n} \e ) \g^{\m\n}
\l^{iA} -\delta \phi^\a {\cal{A}}_\a^A{}_B \l^{iB}
-\frac{1}{\sqrt{2}v_r
c^{r1}} {\cal{A}}_\a^A{}_B \xi^{\a i} \e^B \quad , \nonumber\\
& & \delta \l^{IA}
=-\frac{1}{2\sqrt{2}}\hat{F}^I_{\m\n} \g^{\m\n} \e^A 
- \frac{x^M_r c^{r2}}{2v_s c^{s2}} (\bar{\chi}^M \l^I ) \e^A 
- \frac{x^M_r c^{r2}}{4v_s c^{s2}} (\bar{\chi}^M \e ) \l^{IA}  
\nonumber \\
& & \quad \quad + 
\frac{x^M_r c^{r2}}{8v_s c^{s2}} (\bar{\chi}^M \g_{\m\n} \e ) \g^{\m\n}
\l^{IA} -\delta \phi^\a {\cal{A}}_\a^A{}_B \l^{IB} -\frac{1}{\sqrt{2}v_r
c^{r2}} {\cal{A}}_\a^A{}_B \xi^{\a I} \e^B \quad . 
\eeq
One can compute the commutators of two supersymmetry transformations 
on the bose fields using these relations, and show that they generate 
the local symmetries:
\be
[\delta_1 , \delta_2 ] = \delta_{gct} + \delta_{Lorentz}+\delta_{susy}+
\delta_{tens}+\delta_{gauge} +\delta_{SO(n)}\quad ,
\ee
where the parameters of generic coordinate, local Lorentz, 
supersymmetry,
tensor gauge, vector gauge and composite $SO(n)$ transformations are respectively
\beq
& & \xi_\m = -i (\bar{\e}_1 \g_\m \e_2 ) \quad ,\nonumber \\
& & \W^{mn} = - i \xi^\m (\hat{\w}_\m{}^{mn} - v_r \hat{H}^r_\m{}^{mn} )
-\frac{1}{2}[(\bar{\chi}^M \e_1 )(\bar{\e}_2 \g^{mn} \chi^M )-(\bar{\chi}^M \e_2 )
(\bar{\e}_1 \g^{mn} \chi^M )] \nonumber \\
& & \quad \quad -v_r c^{rz} tr_z [ (\bar{\e}_1 \g^m \l )(\bar{\e}_2 \g^n \l )
- (\bar{\e}_2 \g^m \l )(\bar{\e}_1 \g^n \l )] \quad,\nonumber\\
& & \zeta^A = \xi^\m \psi^A_\m + V^\a_{aC} 
{\cal{A}}_\a^A{}_B \e_2^B (\bar{\e}_1^C \Psi^a ) 
-V^\a_{aC} {\cal{A}}_\a^A{}_B \e_1^B 
(\bar{\e}_2^C \Psi^a )\quad ,\nonumber\\
& & \L^r_\m = -\frac{1}{2}v^r \xi_\m -\xi^\n B^r_{\m\n} \quad , \nonumber\\
& & \L = \xi^\m A_\m\quad, \nonumber\\
& & A^{MN} = \xi^\m x^{Mr} (\de_\m x^N_r )+(\bar{\chi}^M \e_2 )(\bar{\chi}^N \e_1 )
-(\bar{\chi}^M \e_1 )(\bar{\chi}^N \e_2 )\quad .
\eeq
In order to prove this result, one has to use the (anti)self-duality 
condition for the tensor fields, that to all orders in the fermi fields 
is
\be
G_{rs} {\hat{\cal{H}}}^s_{\m\n\r} =\frac{1}{6e}\e_{\m\n\r\sigma\delta\tau}
\hat{\cal{H}}_r^{\sigma\delta\tau} \label{asd}
\ee
in terms of the 3-forms \cite{ns1}
\be
\hat{\cal{H}}^r_{\m\n\r}=\hat{H}^r_{\m\n\r} -\frac{i}{8}v^r (\bar{\chi}^M 
\g_{\m\n\r} \chi^M ) +\frac{i}{8} v^r (\bar{\Psi}_a \g_{\m\n\r} \Psi^a )
-\frac{i}{4} c^{rz} tr_z (\bar{\l} \g_{\m\n\r} \l ) \quad.\label{fs}
\ee
Requiring that the commutator of two supersymmetry transformations 
on the fermi fields close on-shell then 
determines the complete fermi field equations.
The equations  obtained in this way are
\beq & & -i\g^{\m\n\r} D_\n (\hat{\w})\psi_\r^A -\frac{i}{4} v_r 
\hat{H}^r_{\n\sigma\delta}
\g^{\m\n\r}\g^{\sigma\delta} \psi_\r^A -\frac{1}{12}x^M_r \hat{H}^{r 
\n\r\sigma}
\g_{\n\r\sigma} \g^\m
\chi^{MA} \nonumber\\ & & +\frac{1}{2} x^M_r (\hat{\de_\n v^r })\g^\n \g^\m 
\chi^{MA} 
+\frac{3}{2}
\g^{\m\n}\chi^{MA} (\bar{\chi}^M \psi_\n )-\frac{1}{4}
\g^{\m\n} \chi^{MA} (\bar{\chi}^M \g_{\n\r} \psi^\r )\nonumber \\
& &  
+\frac{1}{4} \g_{\n\r} \chi^{MA} (\bar{\chi}^M \g^{\m\n} \psi^\r )
-\frac{1}{2}
\chi^{MA} (\bar{\chi}^M \g^{\m\n} \psi_\n ) +i v_r c^{rz} tr_z
[-\frac{1}{\sqrt{2}}\g^{\n\r}\g^\m \l^A \hat{F}_{\n\r} \nonumber \\ 
& &  + \frac{3i}{4} \g^{\m\n\r}
\l^A (\bar{\psi}_\n
\g_\r \l )-\frac{i}{2} \g^\m \l^A (\bar{\psi}_\n \g^\n \l ) 
+\frac{i}{2}\g^\n 
\l^A (\bar{\psi}_\n
\g^\m \l ) + \frac{i}{4}\g_\r \l^A (\bar{\psi}_\n \g^{\m\n\r}\l )]
\nonumber \\ & &  +\frac{i}{2} x^M_r c^{rz} tr_z [\g_\n \l^A (\bar{\chi}^M 
\g^\n \g^\m \l )]
-V_\a^{aA} \hat{D_\n \phi^\a} \g^\n \g^\m \Psi_a \nonumber \\ & & 
+\frac{i}{\sqrt{2}}[{\cal{A}}_\a^a{}_B \xi^{\a i}\g^\m \l^{iB}+
{\cal{A}}_\a^a{}_B \xi^{\a I}\g^\m \l^{IB}] =0
\eeq
for the gravitino,
\beq 
& & i\g^\m D_\m (\hat{\w})\chi^{M A}-\frac{i}{12}v_r \hat{H}^r_{\m\n\r} \g^{\m\n\r}
\chi^{MA} +\frac{1}{12} x^M_r \hat{H}^r_{ \m\n\r} \g^\sigma \g^{\m\n\r} 
\psi_\sigma^A
+ \frac{1}{2} x^M_r (\hat{\de_\n v^r}) \g^\m \g^\n \psi_\m^A 
\nonumber\\ & & - \frac{1}{\sqrt{2}}x^M_r c^{rz} tr_z (\hat{F}_{\m\n} 
\g^{\m\n} \l^A ) 
+\frac{i}{2} x^M_r c^{rz} tr_z [ \g^\m \g^\n  \l^A (\bar{\psi}_\m \g_\n
\l )] +\frac{1}{2}\g^\m \chi^{NA} (\bar{\chi}^N \g_\m \chi^M ) \nonumber\\ & & +
\frac{3}{8}
v_r c^{rz} tr_z [(\bar{\chi}^M \g_{\m\n} \l )\g^{\m\n}
\l^A ] +\frac{1}{4} v_r c^{rz} tr_z [(\bar{\chi}^M \l ) \l^A ] \nonumber\\ & & +
\frac{3}{2} \frac{x^M_r c^{rz} x^N_s c^{sz}}{v_t c^{tz}} tr_z 
[(\bar{\chi}^N \l ) \l^A ]  -
\frac{1}{4}
\frac{x^M_r c^{rz} x^N_s c^{sz}}{v_t c^{tz}} tr_z [(\bar{\chi}^N \g_{\m\n} \l )
\g^{\m\n} \l^A ]
 \nonumber\\
& & -\frac{x^M_r c^{r1}}{\sqrt{2}v_s c^{s1}}
{\cal{A}}_\a^A{}_B \xi^{\a i} \l^{iB}
-\frac{x^M_r c^{r2}}{\sqrt{2}v_s c^{s2}}
{\cal{A}}_\a^A{}_B \xi^{\a I} \l^{IB}=0
\eeq
for the tensorinos, 
\beq
& & i \g^\m D_\m (\hat{\w})\Psi^a +\frac{i}{12}v_r \hat{H}^r_{\m\n\r}\g^{\m\n\r}
\Psi^a +\g^\m \g^\n \psi_{\m A} V_\a^{aA} \hat{D_\n \phi^\a}+\frac{1}{48} v_r 
c^{rz}tr_z (\bar{\l} \g_{\m\n\r} \l ) \g^{\m\n\r} \Psi^a \nonumber\\
& &  +\frac{1}{12}
\W^{abcd} \g^\m \Psi_b (\bar{\Psi}_c \g_\m \Psi_d )+\sqrt{2}V_\a^{aA}
[\xi^{\a i} \l^i_A + \xi^{\a I} \l^I_A ] =0
\eeq
for the hyperinos. More care is needed in order to derive the equations for the 
gauginos, since the $c_r^z c^{r z^\prime}$ terms in the commutator 
of two supersymmetry transformations are
\beq
& & \frac{c_r^z c^{r z^\prime}}{v_s c^{sz}}
tr_{z^\prime} [\frac{1}{4}(\bar{\e}_1 \g_\m \l^\prime )(\bar{\e}_2
\g_\n \l^\prime ) \g^{\m\n} \l^A  -
\frac{1}{4} (\bar{\l} \g_\m \l^\prime ) (\bar{\e}_1 \g^\m \l^\prime ) \e_2^A -(1
\leftrightarrow 2)
\nonumber\\ & &  - \frac{1}{16} (\bar{\e}_1 \g^\m \e_2 )(\bar{\l}^\prime
\g_{\m\n\r} \l^\prime  )\g^{\n\r}\l^A ]\quad . \label{extra}
\eeq 
If one allows for the term
\be
i  \a c_r^z c^{r z^\prime} tr_{z^\prime} [(\bar{\l} \g_\m \l^\prime )
\g^\m \l^{\prime A}]
\ee
in the gaugino field equation, then what remains of eq. (\ref{extra}) is
\beq
\delta_{extra(\a)} \l^A & & = \frac{c_r^z c^{r
z^\prime}}{v_s c^{sz}} tr_{z^\prime} [\frac{1}{4}(\bar{\e}_1 \g_\m \l^\prime
)(\bar{\e}_2
\g_\n \l^\prime ) \g^{\m\n} \l^A \nonumber\\  
& &+\frac{\a}{2} (\bar{\l} \g_\m \l^\prime
)(\bar{\e}_1 \g_\n \l^\prime )
\g^{\m\n} \e_2^A  -\frac{\a}{16}(\bar{\l}\g_{\m\n\r}\l^\prime )(\bar{\e}_1 \g^\r 
\l^\prime ) \g^{\m\n} \e_2^A \nonumber\\  
& &-\frac{\a}{16} (\bar{\l} \g_\r \l^\prime
)(\bar{\e}_1
\g^{\m\n\r} \l^\prime ) \g_{\m\n} \e_2^A   - \frac{1-\a}{4} (\bar{\l} 
\g_\m \l^\prime )
(\bar{\e}_1
\g^\m \l^\prime ) \e_2^A -(1 \leftrightarrow 2)  
\nonumber\\ 
& & - \frac{1-\a}{16} (\bar{\e}_1 \g^\r \e_2 )(\bar{\l}^\prime
\g_{\m\n\r} \l^\prime  )\g^{\m\n}\l^A ] \quad .\label{central}
\eeq 
As explained in \cite{frs}, no choice of $\a$ can eliminate all these terms, 
that play the role of a central charge felt only by the gauginos. This is the 
``classical'' realization of a general feature: anomalies in current 
conservations are accompanied by related anomalies in current 
commutators \cite{anom}. 
When this is properly taken into account, the field
equations for the gauginos are 
\beq 
& & i v_r c^{rz} \g^\m D_\m (\hat{\w})\l^A +\frac{i}{2} (\hat{\de_\m v_r})c^{rz}
\g^\m \l^A + \frac{i}{2\sqrt{2}}v_r c^{rz} \hat{F}_{\n\r} \g^\m \g^{\n\r} 
\psi_\m^A -
\frac{1}{2\sqrt{2}}x^M_r c^{rz} \hat{F}_{\m\n} \g^{\m\n}\chi^{MA} \nonumber\\ & &
+\frac{i}{12} x^M_r c^{rz} x^M_s \hat{H}^s_{\m\n\r} \g^{\m\n\r}\l^A +
\frac{i}{2} x^M_r c^{rz} (\bar{\chi}^M \l ) \g^\m \psi_\m^A 
 +\frac{i}{4}x^M_r c^{rz} (\bar{\chi}^M \psi_\m )\g^\m \l^A \nonumber\\ & & - 
\frac{i}{8} x^M_r c^{rz} (\bar{\chi}^M \g_{\n\r} \psi_\m ) \g^{\m\n\r} \l^A 
 - \frac{i}{4}x^M_r c^{rz} (\bar{\chi}^M \g_{\m\n}\psi^\m )\g^\n \l^A -
\frac{1}{8}v_r c^{rz} (\bar{\l} \chi^M )\chi^{MA} \nonumber\\ 
& & -\frac{3}{16} v_r c^{rz}
(\bar{\l} \g_{\m\n} \chi^M ) \g^{\m\n} \chi^{MA} -\frac{3}{4} 
\frac{x^M_r c^{rz} x^N_s
c^{sz}}{v_t c^{tz}} (\bar{\l} \chi^M )
\chi^{NA} \nonumber\\ & &  
+\frac{1}{8} \frac{x^M_r c^{rz} x^N_s c^{sz}}{v_t c^{tz}} (\bar{\l} 
\g_{\m\n} \chi^M ) \g^{\m\n} \chi^{NA} 
-\frac{1}{96} (\bar{\Psi}_a \g_{\m\n\r} \Psi^a ) \g^{\m\n\r} \l^A
\nonumber \\ & & 
+ v_r v_s c^{rz} c^{s z^\prime}
tr_{z^\prime} [(\bar{\l} \g_\m \l^\prime )
\g^\m \l^{\prime A} ] - \a c_r^z c^{r z^\prime} tr_{z^\prime} 
[(\bar{\l}\g_\m \l^\prime )
\g^\m \l^{\prime A} ] = 0  \quad .
\label{gauginoeq}
\eeq
Actually to the left-hand side of this equation, valid for the case 
$z\neq 1,2$, one has to add the terms 
\be
- \sqrt{2} V_\a^{aA} \xi^{\a i,I} \Psi_a 
+\frac{i}{\sqrt{2}} {\cal{A}}_\a^A{}_B \xi^{\a i,I } \g^\m \psi_\m^B +
\frac{x^M_r c^{r1,2}}{\sqrt{2} v_s c^{s1,2}} 
{\cal{A}}_\a^A{}_B \xi^{\a i,I} \chi^{MB}
\ee
in the two remaining cases, i.e. for $\l^{i}$ and $\l^{I}$.

Having obtained the complete fermionic field equations, one can add to 
eq. (\ref{lag}) all the terms quartic in the fermi fields, thus obtaining 
the complete lagrangian
\beq 
e^{-1}{\cal{L}} & & =-\frac{1}{4}R +\frac{1}{12}G_{rs} H^{r \m\n\r} H^s_{\m\n\r}
-\frac{1}{4} \de_\m v^r \de^\m v_r + \frac{1}{2} g_{\a\b}(\phi ) 
D_\m \phi^\a D^\m \phi^\b \nonumber\\
& &  +\frac{1}{2} v_r c^{rz} tr_z (F_{\m\n} F^{\m\n}) +\frac{1}{8e}
\e^{\m\n\r\sigma\delta\tau} c_r^z B^r_{\m\n} tr_z (F_{\r\sigma} 
F_{\delta\tau})\nonumber \\
& & 
+\frac{1}{4v_r c^{r1}} {\cal{A}}_\a^A{}_B 
{\cal{A}}_\b^B{}_A \xi^{\a i} \xi^{\b i} 
+\frac{1}{4v_r c^{r2}} {\cal{A}}_\a^A{}_B 
{\cal{A}}_\b^B{}_A \xi^{\a I} \xi^{\b I} 
\nonumber\\
& & -\frac{i}{2}(\bar{\psi}_\m \g^{\m\n\r} D_\n [\frac{1}{2}(\w
+\hat{\w} )]
\psi_\r )-\frac{i}{8}v_r [H+\hat{H}]^{r \m\n\r}(\bar{\psi}_\m \g_\n \psi_\r)
\nonumber \\ 
& & +\frac{i}{48} v_r [H+\hat{H} ]^r_{\r\sigma\delta} (\bar{\psi}_\m
\g^{\m\n\r\sigma\delta}\psi_\n )+\frac{i}{2} (\bar{\chi}^M \g^\m D_\m (\hat{\w})
\chi^M ) \nonumber \\ 
& & -\frac{i}{24}v_r \hat{H}^r_{\m\n\r} (\bar{\chi}^M \g^{\m\n\r}
\chi^M ) +\frac{1}{4}x^M_r [\de_\n v^r +\hat{\de_\n v^r} ](\bar{\psi}_\m \g^\n
\g^\m \chi^M ) \nonumber\\ 
& & -\frac{1}{8} x^M_r [H+\hat{H}]^{r \m\n\r} ( \bar{\psi}_\m
\g_{\n\r}
\chi^M )+\frac{1}{24}x^M_r [H+\hat{H}]^{r \m\n\r} (\bar{\psi}^\sigma 
\g_{\sigma\m\n\r} \chi^M ) \nonumber\\ 
& & +\frac{i}{2} (\bar{\Psi}_a \g^\m D_\m (\hat{\w}) \Psi^a ) + \frac{i}{24}
v_r \hat{H}^r_{\m\n\r} (\bar{\Psi}_a \g^{\m\n\r} \Psi^a ) \nonumber\\
& & - \frac{1}{2} V_\a^{aA}
[ D_\n \phi^\a + \hat{D_\n \phi^\a} ] ( \bar{\psi}_{\m A}\g^{\n} 
\g^{\m} \Psi_{a} )
\nonumber \\
& &  -iv_r c^{rz} tr_z (\bar{\l} \g^\m D_\m (\hat{\w}) \l
)-\frac{i}{12} x^M_r x^M_s \hat{H}^r_{\m\n\r} c^{sz} tr_z (\bar{\l}\g^{\m\n\r} \l )
\nonumber \\ 
& & - \frac{i}{2\sqrt{2}}  v_r c^{rz} tr_z [(F+\hat{F})_{\n\r}(\bar{\psi}_\m
\g^{\n\r} \g^\m \l )] 
 -\frac{1}{\sqrt{2}}x^M_r c^{rz} tr_z [(\bar{\chi}^M \g^{\m\n}
\l )\hat{F}_{\m\n} ] \nonumber \\
& & -\sqrt{2} V_\a^{aA} [\xi^{\a i}(\bar{\l}^i_A \Psi_a ) +\xi^{\a I} (\bar{\l}^I_A
\Psi_a )] +\frac{i}{\sqrt{2}}
{\cal{A}}_\a^A{}_B [\xi^{\a i }(\bar{\l}^i_A \g^\m \psi_\m^B )
+ \xi^{\a I} (\bar{\l}^I_A \g^\m \psi_\m^B )] \nonumber \\
& & +\frac{1}{\sqrt{2}}
{\cal{A}}_\a^A{}_B [ \frac{x^M_r c^{r1}}{v_s c^{s1}}\xi^{\a i}
(\bar{\l}^i_A \chi^{MB} ) + \frac{x^M_r c^{r2}}{v_s c^{s2}}\xi^{\a I}
(\bar{\l}^I_A \chi^{MB} ) ]\nonumber \\
& & +\frac{1}{8}(\bar{\chi}^M \g^{\m\n\r} \chi^M )(\bar{\psi}_\m
\g_\n \psi_\r )-\frac{1}{8}(\bar{\chi}^M \g^\m \chi^N )(\bar{\chi}^M \g_\m \chi^N )
\nonumber \\  
& & +\frac{1}{8} (\bar{\Psi}_a \g^{\m\n\r} \Psi^a )(\psi_\m \g_\n \psi_\r )
+\frac{1}{48} \W^{abcd} (\bar{\Psi}_a \g_\m \Psi_b )(\bar{\Psi}_c \g^\m \Psi_d )
\nonumber \\
& &
 -\frac{1}{16}v_r c^{rz}tr_z (\bar{\l} \g_{\m\n\r} \l )(\bar{\chi}^M
\g^{\m\n\r} \chi^M )  + \frac{i}{8}(\bar{\chi}^M \g_{\m\n}\psi_\r )x^M_r
c^{rz} tr_z (\bar{\l} \g^{\m\n\r} \l )\nonumber \\
& &  + \frac{i}{2} x^M_r c^{rz} tr_z 
[(\bar{\chi}^M \g^\m \g^\n \l ) (\bar{\psi}_\m \g_\n \l )]  -\frac{1}{4} 
(\bar{\psi}_\m \g_\n \psi_\r ) v_r c^{rz}tr_z (\bar{\l} \g^{\m\n\r}
\l )\nonumber\\ 
& & +\frac{1}{8} v_r c^{rz} tr_z [(\bar{\chi}^M \l )(\bar{\chi}^M \l) ]
+ \frac{3}{16}v_r c^{rz}tr_z [(\bar{\chi}^M \g_{\m\n} \l ) (\bar{\chi}^M
\g^{\m\n}
\l )] 
\nonumber\\ 
& & +\frac{3 x^M_r c^{rz}x^N_s c^{sz}}{4 v_t c^{tz}} tr_z
[(\bar{\chi}^M \l )(\bar{\chi}^N \l )] 
 -\frac{x^M_r c^{rz} x^N_s c^{sz}}{8 v_t c^{tz}} tr_z [(\bar{\chi}^M 
\g_{\m\n} \l )(\bar{\chi}^N \g^{\m\n}\l )] \nonumber\\ 
& & 
-\frac{5}{96} v_r c^{rz} tr_z (\bar{\l} \g_{\m\n\r}\l )(\bar{\Psi}_a 
\g^{\m\n\r}\Psi^a )
 -\frac{1}{2} v_r v_s c^{rz}c^{s z^\prime} tr_{z,z^\prime} [(\bar{\l}\g_\m \l^\prime
)(\bar{\l} \g^\m \l^\prime ) ] \nonumber\\ 
& & +\frac{\a}{2}c^{rz} c_r^{z^\prime} tr_{z,z^\prime} [(\bar{\l} \g_\m
\l^\prime )(\bar{\l} \g^\m \l^\prime )] \quad .
\label{completelag}
\eeq
From this lagrangian, in the 1.5 order formalism and using the (anti)self-duality 
conditions of eqs. (\ref{asd}) and (\ref{fs}), one can obtain the remaining
complete bosonic field equations. 
Once more, it is important to notice that this lagrangian in neither gauge 
invariant nor supersymmetric: its variation under gauge transformations produces 
the gauge anomaly of eq. (\ref{consanomaly}), while its variation under 
the complete supersymmetry transformations produces the complete 
supersymmetry anomaly
\beq {\cal{A}}_\e & &=c_r^z c^{r z^\prime} tr_{z, z^\prime} \lbrace -\frac{1}{4}
\e^{\m\n\r\sigma\delta\tau}\delta_\e A_\m A_\n F^\prime_{\r\sigma} 
F^\prime_{\delta\tau} -\frac{1}{6}
\e^{\m\n\r\sigma\delta\tau} \delta_\e A_\m F_{\n\r} 
\w^\prime_{\sigma\delta\tau} \nonumber\\ 
& & +\frac{i e}{2} \delta_\e A_\m F_{\n\r}
(\bar{\l}^\prime
\g^{\m\n\r} 
\l^\prime )+\frac{i e}{2} \delta_\e A_\m (\bar{\l} \g^{\m\n\r} \l^\prime )
F^\prime_{\n\r}  + ie\delta_\e A_\m (\bar{\l}\g_\n \l^\prime ) F^{\prime \m\n}
\nonumber\\ 
& & +
\frac{e}{32} \delta_\e e_\m{}^m (\bar{\l} \g^{\m\n\r} \l )(\bar{\l}^\prime
\g_{m \n\r} \l^\prime )  -\frac{e}{2\sqrt{2}} \delta_\e 
A_\m (\bar{\l} \g^\m \g^\n \g^\r 
\l^\prime )(\bar{\l}^\prime \g_\n \psi_\r )  \nonumber\\ 
& & + \frac{e x^M_s c^{s
z^\prime}}{v_t c^{t z^\prime}} [-\frac{3 i}{2\sqrt{2}}
 \delta_\e A_\m (\bar{\l} \g^\m \l^\prime )(\bar{\l}^\prime \chi^M )
 -\frac{i}{4 \sqrt{2}} \delta_\e A_\m (\bar{\l} \g^{\m\n\r} \l^\prime )
(\bar{\l}^\prime \g_{\n\r}\chi^M ) \nonumber\\ 
& & - \frac{i}{2\sqrt{2}} \delta_\e A_\m  (\bar{\l} \g_\n
\l^\prime )(\bar{\l}^\prime \g^{\m\n} \chi^M )] +\frac{\a}{2} \delta_\e
[ e (\bar{\l} \g_\m \l^\prime )(\bar{\l} \g^\m \l^\prime ) ] \rbrace 
\nonumber \\
& & +\frac{i e c_r^1 c^{rz}}{2 v_s c^{s 1}}
{\cal{A}}_\a^A{}_B \xi^{\a i}  tr_z [\delta_\e A_\m
(\bar{\l}^i_A \g^\m \l^B )] \nonumber\\
& & +\frac{i e c_r^2 c^{rz}}{2 v_s c^{s 2}} 
{\cal{A}}_\a^A{}_B \xi^{\a I}tr_z [\delta_\e A_\m (\bar{\l}^I_A
\g^\m \l^B )]
\ .
\label{theanomaly}
\eeq

The presence of a term proportional to the parameter $\a$ in eq. 
(\ref{completelag}) reflects the general fact that anomalies are defined up to 
the variation of a local functional. 
Gauge and supersymmetry anomalies are in general related by the Wess-Zumino 
consistency conditions \cite{wz}
\beq
& & \delta_\e {\cal{A}}_\L = \delta_\L {\cal{A}}_\e \quad , \nonumber \\
& & \delta_{\e_1} {\cal{A}}_{\e_2} -  \delta_{\e_2} {\cal{A}}_{\e_1}=
{\cal{A}}_\L + {\cal{A}}_\zeta \quad .
\eeq
What is peculiar of these six-dimensional models is the fact that the second 
condition closes only on-shell, and precisely on the gaugino field equations 
\cite{frs}. 
Since the inclusion of the term proportional to $\a$ in the lagrangian 
modifies both these 
equations and the supersymmetry anomaly, there must be some extra terms
that permit the Wess-Zumino conditions to close on-shell 
for every value of $\a$.    
This is precisely the role of the terms in eq. (\ref{central}) in the 
commutator of two supersymmetry transformations on the gauginos, that thus can 
be seen as a transformation needed in order to close the Wess-Zumino conditions
precisely on the field equations determined by the algebra. Since the Wess-Zumino 
conditions need only the equation of the gauginos, only these fields sense the 
additional transformation (the whole construction is explained in more detail in 
\cite{frs}).

\section{Inclusion of abelian vectors} 
Up to now, we have always considered the case in which the gauge group is 
non-abelian. In the abelian case, the couplings can actually 
have a more general form, since
gauge invariance allows non-diagonal kinetic and Chern-Simons terms, 
in which the constants $c^r_z$ are substituted by generic symmetric matrices
$c^r_{IJ}$, with $I,J$ running over the various $U(1)$ factors \cite{cjlp}. 
In \cite{r}, the complete coupling of minimal 
six-dimensional supergravity to tensor multiplets and abelian vector multiplets 
was constructed. We now want to generalize it to the case in which also 
charged hypermultiplets are present, and therefore we will consider the gauging 
with respect to abelian subgroups of $USp(2)\times USp(2n_H )$. There are no 
subtleties when the symmetric matrices $c^r_{IJ}$ are diagonal (or 
simultaneously diagonalizable), since in this situation the results of the 
previous section can be straightforwardly applied. We are thus interested in 
the case in which the $c^r_{IJ}$ can not be simultaneously diagonalized.  
To this end, we will consider a model in which only 
these abelian gauge groups are present. 
The most general situation can be obtained  
combining the following results with those obtained in the previous section.

We denote with $A_\m^I$, $I=1,...,m$, the set of abelian vectors, and
the gauginos are correspondingly denoted by $\l^{IA}$. 
We collect here only the final results, since the construction 
follows the same lines 
as in the non-abelian case. All the field equations may then be derived from the 
lagrangian
\beq 
e^{-1}{\cal{L}} & & =-\frac{1}{4}R +\frac{1}{12}G_{rs} H^{r \m\n\r} H^s_{\m\n\r}
-\frac{1}{4} \de_\m v^r \de^\m v_r \nonumber\\
& &  -\frac{1}{4} v_r c^{rIJ} F_{\m\n}^I F^{J \m\n} - \frac{1}{16e}
\e^{\m\n\r\sigma\delta\tau} c_r^{IJ} B^r_{\m\n} F_{\r\sigma}^I 
F_{\delta\tau}^J \nonumber\\
& &  + \frac{1}{2} g_{\a\b}(\phi ) D_\m \phi^\a D^\m \phi^\b 
+\frac{1}{4}[(v \cdot c )^{-1}]^{IJ} 
{\cal{A}}_\a^A{}_B {\cal{A}}_\b^B{}_A \xi^{\a I} \xi^{\b J}
\nonumber\\
& & -\frac{i}{2}(\bar{\psi}_\m \g^{\m\n\r} D_\n [\frac{1}{2}(\w
+\hat{\w} )]
\psi_\r )-\frac{i}{8}v_r [H+\hat{H}]^{r \m\n\r}(\bar{\psi}_\m \g_\n \psi_\r)
\nonumber \\ 
& & +\frac{i}{48} v_r [H+\hat{H} ]^r_{\r\sigma\delta} (\bar{\psi}_\m
\g^{\m\n\r\sigma\delta}\psi_\n )+\frac{i}{2} (\bar{\chi}^M \g^\m D_\m (\hat{\w})
\chi^M ) \nonumber \\ 
& & -\frac{i}{24}v_r \hat{H}^r_{\m\n\r} (\bar{\chi}^M \g^{\m\n\r}
\chi^M ) +\frac{1}{4}x^M_r [\de_\n v^r +\hat{\de_\n v^r} ](\bar{\psi}_\m \g^\n
\g^\m \chi^M ) \nonumber\\ 
& & -\frac{1}{8} x^M_r [H+\hat{H}]^{r \m\n\r} ( \bar{\psi}_\m
\g_{\n\r}
\chi^M )+\frac{1}{24}x^M_r [H+\hat{H}]^{r \m\n\r} (\bar{\psi}^\sigma 
\g_{\sigma\m\n\r} \chi^M ) \nonumber\\ 
& & +\frac{i}{2} (\bar{\Psi}_a \g^\m D_\m (\hat{\w}) \Psi^a ) + \frac{i}{24}
v_r \hat{H}^r_{\m\n\r} (\bar{\Psi}_a \g^{\m\n\r} \Psi^a ) \nonumber\\
& & - \frac{1}{2} V_\a^{aA}
[ D_\n \phi^\a + \hat{D_\n \phi^\a} ] (\bar{\psi}_{\m A} \g^{\n} \g^{\m}
\Psi_{a} )\nonumber \\
& &  +\frac{i}{2}v_r c^{rIJ} (\bar{\l}^I \g^\m D_\m (\hat{\w}) \l^J
)+\frac{i}{24} x^M_r x^M_s \hat{H}^r_{\m\n\r} c^{sIJ} (\bar{\l}^I\g^{\m\n\r} \l^J )
\nonumber \\ 
& & + \frac{i}{4\sqrt{2}}  v_r c^{rIJ} (F+\hat{F})_{\n\r}^I (\bar{\psi}_\m
\g^{\n\r} \g^\m \l^J ) 
 +\frac{1}{2\sqrt{2}}x^M_r c^{rIJ} (\bar{\chi}^M \g^{\m\n}
\l^I )\hat{F}_{\m\n}^J  \nonumber\\ 
& & -\sqrt{2} V_\a^{aA} \xi^{\a I} (\bar{\l}^I_A
\Psi_a ) +\frac{i}{\sqrt{2}} {\cal{A}}_\a^A{}_B 
\xi^{\a I} (\bar{\l}^I_A \g^\m \psi_\m^B )\nonumber \\
& &   +\frac{1}{\sqrt{2}}[(v \cdot c )^{-1} (x^M \cdot c )]^{IJ}
{\cal{A}}_\a^A{}_B \xi^{\a I} ( \bar{\l}^J_A \chi^{MB} ) \nonumber \\
& & +\frac{1}{8}(\bar{\chi}^M \g^{\m\n\r} \chi^M )(\bar{\psi}_\m
\g_\n \psi_\r )-\frac{1}{8}(\bar{\chi}^M \g^\m \chi^N )(\bar{\chi}^M \g_\m \chi^N )
\nonumber \\  
& & +\frac{1}{8} (\bar{\Psi}_a \g^{\m\n\r} \Psi^a )(\psi_\m \g_\n \psi_\r )
+\frac{1}{48} \W^{abcd} (\bar{\Psi}_a \g_\m \Psi_{b} )(\bar{\Psi}_c 
\g^\m \Psi_{d} )
\nonumber \\
& &
 +\frac{1}{32}v_r c^{rIJ} (\bar{\l}^I \g_{\m\n\r} \l^J )(\bar{\chi}^M
\g^{\m\n\r} \chi^M )  - \frac{i}{16}(\bar{\chi}^M \g_{\m\n}\psi_\r )x^M_r
c^{rIJ} (\bar{\l}^I \g^{\m\n\r} \l^J )\nonumber \\
& &  -\frac{i}{4} x^M_r c^{rIJ} 
(\bar{\chi}^M \g^\m \g^\n \l^I ) (\bar{\psi}_\m \g_\n \l^J )  
+\frac{1}{8} 
(\bar{\psi}_\m \g_\n \psi_\r ) v_r c^{rIJ}(\bar{\l}^I \g^{\m\n\r}
\l^J )\nonumber\\ 
& &  -\frac{1}{16} v_r c^{rIJ}(\bar{\chi}^M \l^I )(\bar{\chi}^M \l^J ) 
- \frac{3}{32}v_r c^{rIJ} (\bar{\chi}^M \g_{\m\n} \l^I ) (\bar{\chi}^M
\g^{\m\n} \l^J ) \nonumber\\ 
& & +[ (x^M \cdot c )(v \cdot c )^{-1} (x^N \cdot c )]^{IJ}
[-\frac{1}{4}(\bar{\chi}^M \l^I )(\bar{\chi}^N \l^J ) \nonumber \\
& & +\frac{1}{16} (\bar{\chi}^N
\g^{\m\n} \l^I )(\bar{\chi}^M \g_{\m\n} \l^J ) -\frac{1}{8}(\bar{\chi}^N \l^I )
(\bar{\chi}^M \l^J )]
\nonumber\\ 
& & 
+\frac{5}{192} v_r c^{rIJ} (\bar{\l}^I \g_{\m\n\r}\l^J )(\bar{\Psi}_a 
\g^{\m\n\r}\Psi^a )
 -\frac{1}{8} v_r v_s c^{rIJ}c^{s KL} (\bar{\l}^I \g_\m \l^K
)(\bar{\l}^J \g^\m \l^L )  \nonumber\\ 
& & +\frac{\a}{8}c^{rIJ} c_r^{KL} (\bar{\l}^I \g_\m
\l^K )(\bar{\l}^J \g^\m \l^L )] \quad .
\eeq
The variation of this lagrangian with respect to gauge 
transformations gives the
abelian gauge anomaly
\be
{\cal{A}}_{\L}=-\frac{1}{32}\e^{\m\n\r\sigma\d\tau}c_{r}^{IJ}c^{rKL}\L^{I}F^{J}_{\m\n}
F^{K}_{\r\sigma}  F^{L}_{\d\tau}\quad , \label{abeliananomaly}  
\ee
while its variation with respect to the supersymmetry transformations 
\beq 
& & \delta e_\m{}^m = -i ( \bar{\e} \g^m \psi_\m ) \quad , \nonumber\\ 
& & \delta B^r_{\m\n} =i v^r ( \bar{\psi}_{[\m} \g_{\n]} \e )
+\frac{1}{2} x^{Mr} ( \bar{\chi}^M \g_{\m\n} \e ) 
+ 2c^{rIJ}  A_{[\m}^I \delta A_{\n]}^J 
\quad , \nonumber\\ 
& & \delta v_r = x^M_r ( \bar{\e} \chi^M ) \quad , \qquad \delta x^M_r = v_r
(\bar{\e} \chi^M )\quad , \nonumber\\
& & \delta \phi^\a = V^\a_{aA} ({\bar{\e}}^A \Psi^a ) \quad ,\nonumber\\
& & \delta A_\m^I = -\frac{i}{\sqrt{2}} (\bar{\e} \g_\m \l^I )\quad, 
\nonumber\\
& & \delta \psi_\m^A =D_\m (\hat{\w})\e^A +\frac{1}{4} v_r \hat{H}^r_{\m\n\r}
\g^{\n\r}\e^{A} -\frac{3i}{8} \g_\m \chi^{MA} (\bar{\e} \chi^M ) -\frac{i}{8} 
\g^\n \chi^{MA}
(\bar{\e} \g_{\m\n} \chi^M )\nonumber\\
& & \quad \quad +\frac{i}{16} \g_{\m\n\r} \chi^{MA} (\bar{\e} 
\g^{\n\r} \chi^M ) 
- \frac{9i}{16} v_r c^{rIJ} \l^{IA} (\bar{\e} \g_\m \l^J) + 
\frac{i}{16} v_r c^{rIJ} \g_{\m\n} \l^{IA} (\bar{\e} \g^\n \l^J )\nonumber\\
& & \quad \quad - \frac{i}{32} v_r c^{rIJ} \g^{\n\r} \l^{IA} (\bar{\e}
\g_{\m\n\r} \l^J ) -\delta \phi^\a 
{\cal{A}}_\a^A{}_B \psi_\m^B \quad ,\nonumber\\ 
& & \delta \chi^{MA} =
\frac{i}{2} x^M_r (\hat{\de_\m v^r} ) \g^\m \e^A +
\frac{i}{12} x^M_r \hat{H}^r_{\m\n\r} \g^{\m\n\r}\e^A \nonumber\\
& & \quad \quad +\frac{1}{4} x^M_r c^{rIJ}  \g_\m \l^{IA} (\bar{\e} \g^\m \l^J )  
-\delta \phi^\a {\cal{A}}_\a^A{}_B \chi^{MB} \quad ,\nonumber\\ 
& & \delta \Psi^a = i \g^\m \e_A V_\a^{aA} \hat{D_\m \phi^\a} -\delta \phi^\a
{\cal{A}}_\a^a{}_b \Psi^b \quad , \nonumber \\
& & \delta \l^{IA}
=-\frac{1}{2\sqrt{2}}\hat{F}_{\m\n}^I \g^{\m\n} \e^A +[(v \cdot c)^{-1} (x^M \cdot
c )]^{IJ}[
- \frac{1}{2} (\bar{\chi}^M \l^J ) \e^A 
- \frac{1}{4} (\bar{\chi}^M \e ) \l^{JA}  
\nonumber \\
& & \quad \quad + 
\frac{1}{8} (\bar{\chi}^M \g_{\m\n} \e ) \g^{\m\n}
\l^{JA}] -\delta \phi^\a 
{\cal{A}}_\a^A{}_B \l^{IB} -\frac{1}{\sqrt{2}}[(v \cdot c)^{-1}
]^{IJ} {\cal{A}}_\a^A{}_B \xi^{\a J } \e^B
\eeq
gives the supersymmetry anomaly
\beq {\cal{A}}_\e & &=c_r^{IJ} c^{r KL}  \lbrace -\frac{1}{16}
\e^{\m\n\r\sigma\delta\tau}\delta_\e A_\m^I A_\n^J F^K_{\r\sigma} 
F^L_{\delta\tau} -\frac{1}{8}
\e^{\m\n\r\sigma\delta\tau} \delta_\e A_\m^I F_{\n\r}^J A_\sigma^K F_{\delta\tau}^L
\nonumber\\ 
& & +\frac{i e}{8} \delta_\e A_\m^I F_{\n\r}^J
(\bar{\l}^K
\g^{\m\n\r} 
\l^L )+\frac{i e}{8} \delta_\e A_\m^I (\bar{\l}^J \g^{\m\n\r} \l^K )
F^L_{\n\r}  + \frac{ie}{4}\delta_\e A_\m^I (\bar{\l}^J
\g_\n \l^K ) F^{L \m\n}
\nonumber\\ 
& & 
+\frac{e}{128} \delta_\e e_\m{}^m (\bar{\l}^I \g^{\m\n\r} \l^J )(\bar{\l}^K
\g_{m \n\r} \l^L )  -\frac{e}{8\sqrt{2}} \delta_\e A_\m^I (\bar{\l}^J \g^\m \g^\n 
\g^\r 
\l^K )(\bar{\l}^L \g_\n \psi_\r ) \rbrace \nonumber\\ 
& & + e c_r^{IJ}[ c^r (v \cdot c )^{-1} (x^M \cdot c)]^{KL}
\lbrace -\frac{i}{4\sqrt{2}}\delta_\e A_\m^I (\bar{\l}^J \g^\m \l^K )(
\bar{\chi}^M \l^L )\nonumber \\
& & +\frac{i}{16\sqrt{2}} \delta_\e A_\m^I (\bar{\l}^J
\g^\m \g^{\n\r} \l^L )(\bar{\chi}^M \g_{\n\r} \l^K ) -\frac{i}{8 \sqrt{2}}
\delta_\e A_\m^I (\bar{\l}^J \g^\m \l^L )(\bar{\chi}^M \l^K ) \rbrace\nonumber\\
& & -\frac{ie}{4}c_r^{IJ} [(v \cdot c )^{-1} c^r ]^{KL}
\delta_\e A_\m^I {\cal{A}}_\a^A{}_B \xi^{\a K} (\bar{\l}^L_A \g^\m \l^{JB} )
\nonumber\\
& & +\frac{\a}{8} c_r^{IJ} c^{rKL}\delta_\e
[ e (\bar{\l}^I \g_\m \l^K )(\bar{\l}^J \g^\m \l^L ) ]  \quad .
\eeq
Once again, in the case of the gauginos, aside from 
local symmetry transformations and field equations,
the commutator of two supersymmetry transformations generates the additional 
two-cocycle 
\beq
 \delta_{(\a)} \l^I & &=
[(v\cdot c )^{-1}c_r ]^{IJ}c^{rKL}  [-\frac{1}{8}(\bar{\e}_1
\g_\m \l^K )(\bar{\e}_2
\g_\n \l^L ) \g^{\m\n} \l^J -\frac{\a}{4} (\bar{\l}^J \g_\m
\l^K )(\bar{\e}_1 \g_\n \l^L )
\g^{\m\n} \e_2 \nonumber\\  
& & +\frac{\a}{32}(\bar{\l}^J \g_{\m\n\r}\l^K )(\bar{\e}_1 \g^\r 
\l^L ) \g^{\m\n} \e_2 +\frac{\a}{32} (\bar{\l}^J \g_\r
\l^K )(\bar{\e}_1
\g^{\m\n\r} \l^L ) \g_{\m\n} \e_2  \nonumber\\  
& & + \frac{1-\a}{8} (\bar{\l}^J \g_\m
\l^K ) (\bar{\e}_1
\g^\m \l^L ) \e_2 -(1 \leftrightarrow 2)  \nonumber\\
& & 
+ \frac{1-\a}{32} (\bar{\e}_1 \g^\m \e_2 )(\bar{\l}^K
\g_{\m\n\r} \l^L  )\g^{\n\r}\l^J ] \quad .\label{twococycle}
\eeq 
All the observations made for the non-abelian case are naturally valid also 
here: the theory is obtained by the requirement that the Wess-Zumino
conditions close on-shell, and, as we have already shown, it is determined 
up to an arbitrary quartic coupling for the gauginos. 
In the case of a single vector 
multiplet, in which this quartic coupling
vanishes, the two-cocycle of eq. (\ref{twococycle}) is still present, 
although it is properly independent of $\a$. The tensionless string 
phase transition point in the moduli space of the scalars in the tensor multiplets
now would correspond to the vanishing of some of the 
eigenvalues of the matrix $(v \cdot c )^{IJ}$
\cite{r}. 

\section{Discussion}
In the previous sections we have completed the coupling of (1,0) six-dimensional 
supergravity to tensor multiplets, vector multiplets and 
charged hypermultiplets. We have derived all the field equations from the 
lagrangian (\ref{completelag}), with the prescription that the 
(anti)self-duality 
conditions of eq. (\ref{asd}) must be used after varying. Moreover, 
the variation 
of the lagrangian with respect to the 2-forms gives the divergence of the 
(anti)self-duality conditions. 
We want now to apply to our case the general method introduced by
Pasti, Sorokin and Tonin \cite{pst} for obtaining Lorentz-covariant 
lagrangians for (anti)self-dual tensors using a single auxiliary field. 
Alternative constructions \cite{infinite},
some of which preceded the work of PST, need an infinite number of
auxiliary fields, and bear a closer relationship to the BRST formulation
of closed-string spectra \cite{berkovits}. 
This method has already been applied to a number of systems, including (1,0) 
six-dimensional supergravity coupled to tensor multiplets \cite{dlt}, type IIB 
ten-dimensional supergravity \cite{dlt2} and (1,0) six-dimensional supergravity 
coupled to vector and tensor multiplets \cite{rs2,r}.

Our theory describes a single self-dual 3-form
\be
\hat{\cal{H}}_{\m\n\r}=v_r \hat{H}^r_{\m\n\r} -\frac{i}{8}(\bar{\chi}^M \g_{\m\n\r}
\chi^M )+ \frac{i}{8}(\bar{\Psi}_a \g_{\m\n\r}\Psi^a ) 
\ee
and $n_T$ antiself-dual 3-forms
\be
\hat{\cal{H}}^M_{\m\n\r}= x^M_r \hat{H}^r_{\m\n\r} -\frac{i}{4}x^M_r 
c^{rz} tr_z (\bar{\l} \g_{\m\n\r} \l ) \quad .
\ee
The complete Lagrangian is obtained adding to eq. (\ref{completelag}) the term
\be
-\frac{\de^\m \phi \de^\s \phi}{4 (\de \phi )^2}[\hat{\cal{H}}^-_{\m\n\r}
\hat{\cal{H}}^-_\s{}^{\n\r} +\hat{\cal{H}}^{M+}_{\m\n\r} 
\hat{\cal{H}}^{M+}{}_\s{}^{\n\r} ]
\quad ,
\ee
where $\phi$ is an auxiliary field and $H^{\pm} = H \pm * H$. 
The resulting lagrangian is invariant under the additional gauge transformations
\cite{pst}
\be
\delta B^r_{\m\n}= (\de_\m \phi )\L^r_\n - (\de_\n \phi )\L^r_\m
\ee
and
\be
\delta \phi =\L \quad , \qquad \delta B^r_{\m\n} =\frac{\L}{(\de \phi )^2}
[v^r \hat{\cal{H}}^-_{\m\n\r} -x^{Mr}\hat{\cal{H}}^{M+}_{\m\n\r}]\de^\r \phi \quad ,
\ee
used to recover the usual field equations for (anti)self-dual forms.
The 3-form
\be
\hat{K}_{\m\n\r} =\hat{\cal{H}}_{\m\n\r}-
3\frac{\de_{[\m} \phi \de^\s \phi }{(\de \phi )^2 }\hat{\cal{H}}^-_{\n\r ]\s}
\ee
is {\it identically} self-dual, while the 3-forms
\be
\hat{K}^M_{\m\n\r} =\hat{\cal{H}}^M_{\m\n\r}-
3\frac{\de_{[\m} \phi \de^\s \phi }{(\de \phi )^2 }\hat{\cal{H}}^{M +}_{\n\r ]\s}
\ee
are  {\it identically} antiself-dual \cite{dlt}. In order to obtain the complete 
supersymmetry transformations, we have to substitute $\hat{\cal{H}}$ with
$\hat{K}$ in the transformation  of the gravitino and $\hat{\cal{H}}^M$ with
$\hat{K}^M$ in the transformations of the tensorinos. Moreover, the auxiliary scalar
is invariant under supersymmetry \cite{dlt,dlt2}. 
It can be shown that the complete lagrangian transforms under supersymmetry 
as dictated by the Wess-Zumino consistency conditions.
The commutator of two 
supersymmetry transformations on $B^r_{\m\n}$ 
now generates the local PST transformations
with parameters
\be
\L_{r \m} =\frac{\de^\s \phi}{(\de \phi )^2 }
(v_r \hat{\cal H}^-_{\s\m\n }-x^M_r \hat{\cal H}^{M+}_{\s\m\n} )\xi^\n 
\quad , \qquad
\L = \xi^\m \de_\m \phi \quad ,
\ee
while in the parameter of the local Lorentz transformation the term 
$\hat{\cal{H}}$ is replaced by $\hat{K}$. All other parameters 
remain unchanged. 

It would be interesting to study in some detail the vacua of the 
lagrangian (\ref{completelag}), analyzing the extrema of the potential 
(\ref{pot}). 
As a simple example, consider the model without
hypermultiplets, in which one can gauge the global 
R-symmetry group $USp(2)$ of the theory. Formally, the gauged theory 
without hypermultiplets is obtained from the theory described previously
putting $n_H=0$ and making the identification
\be
{\cal{A}}_\a^A{}_B \xi^{\a i} \rightarrow - T^{iA}{}_B \quad,
\ee
where $T^i$ are the anti-hermitian generators of $USp(2)$. This 
corresponds to the replacement of the previous 
couplings between gauge fields and spinors, 
dressed by the scalars in the case $n_H \neq 0$, with ordinary minimal couplings:
\be
D_\m \phi^\a {\cal{A}}_\a^A{}_B \e^B \rightarrow A_\m^A{}_B \e^B \quad .
\ee
Implementing this identification gives in this case the positive-definite
potential  
\be
V =\frac{3}{8 v_r c^{r1}} 
\ee
for the scalars in the tensor multiplets. 
One would thus expect that in these models supersymmetry be spontaneously broken. 
Notice that this potential diverges at the tensionless string phase transition 
point. 
Similarly, one could  try to study explicitly the behavior of the potential 
in simple models containing charged hypermultiplets. 
Their dimensional reduction  gives N=2 supergravity
coupled to vector and hypermultiplets in five dimensions, and 
in the context of the AdS/CFT
correspondence  and its generalizations \cite{m}
there is a renewed interest in studying the explicit
gauging  of these five-dimensional models (see, for instance, 
\cite{bc} and references therein). Notice that in five dimensions
the anomaly that results from
the dimensional reduction of our model  can be canceled by a 
local counterterm, and thus the low-energy effective action does not 
present the subtleties of the six-dimensional case 
\cite{fms}.

The couplings we have derived here are the most general couplings of 
$(1,0)$ six-dimensional supergravity to vector, tensor and hypermultiplets. 
One may wonder if one had the option to gauge a subgroup of $SO(1,n_T)$, 
the isometry group of the scalars in the tensor multiplets. Of course, 
we do not know 
how to write a gauge covariant field-strength for antisymmetric tensor fields, 
but there is a more direct reason why this gauging is not expected  
to work, namely
the fact that once we couple vector and tensor multiplets, the $SO(1,n_T)$ 
transformations are no longer  global symmetries of the theory, because of 
the presence of the matrices $c^r$. 

\vskip 24pt
\section*{Acknowledgments}
I am grateful to C. Angelantonj, S. Ferrara and  A. Sagnotti 
for interesting and helpful discussions. I also thank the organizers of 
the ``Semestre on Supergravity, Superstrings and M-Theory'' and all 
the members of the Centre Emile Borel for the hospitality during the 
period in which this work has been completed. 
This work has been supported in part by a Marie Curie Fellowship of the European 
Community Programme ``Improving human research potential and the socio-economic 
knowledge base'' under contract number HPMT-CT-2000-00165, in part 
by the EEC contract HPRN-CT-2000-00122 and in part by the INTAS project 991590.
\vskip 36pt

\section*{Appendix}
The conventions used in this paper are similar to those of \cite{frs}.
The indices of $USp(2)$ and $USp(2 n_H)$ are raised and lowered by the 
antisymmetric symplectic invariant tensors $\e^{AB}$ and $\W^{ab}$ with 
the following conventions:
$$
V^A = \e^{AB} V_B \quad , 
\qquad V_A = \e_{BA} V^B \quad  \qquad (\e^{AB} \e_{AC} =\d^B_C )\quad ,
$$
$$
W^a = \W^{ab} W_b \quad , \qquad W_a = \W_{ba} W^b \quad \qquad
( \W^{ab} \W_{ac}  = \d^b_c )\quad .
$$
All spinors satisfy symplectic Majorana-Weyl conditions. In particular, spinors 
with  $USp(2)$ indices satisfy the condition
$$
\Psi^A = \e^{AB} C \bar{\Psi}^T_B \quad,
$$
while spinors with $USp(2 n_H)$ indices satisfy the condition
$$
\Psi^a = \W^{ab} C \bar{\Psi}^T_b \quad,
$$
where
$$
\bar{\Psi}_{A,a} = (\Psi^{a,A} )^{\dag} \g_0 \quad .
$$
From these relations one can deduce the properties of spinor bilinears under 
Majorana flip. For instance:
$$
(\bar{\chi}_A \Psi^a ) = \e_{AB} \W^{ab} (\bar{\Psi}_b \chi^B ) \quad ,
$$
and similar relations when $\g$-matrices are included.
In our notations a spinor bilinear with two $USp(2)$ indices contracted is 
written without explicit indices, {\it i.e.}
$$
(\bar{\chi}_A \Psi^A ) \equiv (\bar{\chi} \Psi ) \quad,
$$
while in all the other bilinears the symplectic indices are explicit.

The connections ${\cal{A}}_\a^A{}_B$ and ${\cal{A}}_\a^a{}_b$ are anti-hermitian.
Belonging to the adjoint representation of a symplectic group, 
they are symmetric if considered 
with both upper or both lower indices. 

The anti-hermitian generators $T^i$ and $T^I$ satisfy 
the commutation relations
$$
[ T^i , T^j ] =\e^{ijk} T^k \quad , \qquad [ T^I , T^J ] =f^{IJK} T^K \quad ,
$$
as well as the trace conditions
$$
tr ( T^i T^j ) = -\frac{1}{2} \d^{ij} \quad ,\qquad  
tr ( T^I T^J ) = -\frac{1}{2} \d^{IJ} \quad .
$$

\end{document}